\documentclass[conference]{IEEEtran}
\pdfoutput=1
\usepackage{tikz}
\usepackage{amsfonts}

\usepackage{tikz}
\usetikzlibrary{arrows,decorations.pathmorphing,backgrounds,positioning,fit,matrix} 

\IEEEoverridecommandlockouts

\title{Arithmetic Distribution Matching}

\IEEEoverridecommandlockouts

\author{\IEEEauthorblockN{Sebastian Baur and Georg B\"ocherer}
\IEEEauthorblockA{Institute for Communications Engineering\\Technische Universit\"at M\"unchen, Germany\\
Email: \texttt{baursebastian@mytum.de,georg.boecherer@tum.de}}
\thanks{This work was supported by the German Ministry of Education and Research in the framework of an Alexander von Humboldt Professorship.}
}

\usepackage{amsmath}
\usepackage{amsthm}
\usepackage{amssymb}
\usepackage{bm}
\usepackage{xspace}
\usepackage{xcolor}
\usepackage{graphicx}
\usepackage{url}
\usepackage{framed}
\usepackage{float}
\usepackage{rotating}
\usepackage{verbatim}
\usepackage{listings}
\usepackage{algorithm}
\usepackage[noend]{algpseudocode}

\usepackage{lscape}

\usepackage{pgfplots}
\usepackage{pgf}
\usepackage{tikz}
\usetikzlibrary{arrows,shapes.misc,chains,scopes}
\usepackage{pgfplotstable}
\usepackage{booktabs}
\usepackage{colortbl}

\newcommand{\executeiffilenewer}[3]{%
\ifnum\pdfstrcmp{\pdffilemoddate{#1}}%
{\pdffilemoddate{#2}}>0%
{\immediate\write18{#3}}\fi%
}

\graphicspath{{figures/}}

\theoremstyle{plain}
\newtheorem{proposition}{Proposition}%[chapter]

\newcounter{examplecount}

\newcommand{\vecc}{\boldsymbol{c}}

\newcommand{\vecs}{\boldsymbol{s}}

\newcommand{\setc}{\ensuremath{\mathcal{C}}\xspace}

\newcommand{\bmm}{\begin{matrix}}
\newcommand{\emm}{\end{matrix}}
\newcommand{\bpm}{\begin{pmatrix}}
\newcommand{\epm}{\end{pmatrix}}

\newcommand{\bsbm}{\left[\begin{smallmatrix}}
\newcommand{\esbm}{\end{smallmatrix}\right]}

\newcommand{\bbm}{\begin{bmatrix}}
\newcommand{\ebm}{\end{bmatrix}}

\DeclareMathOperator{\expop}{\mathbb{E}}
\DeclareMathOperator{\entop}{\mathbb{H}}

\DeclareMathOperator{\kl}{\mathbb{D}}

\newcommand{\oeq}[1]{\overset{(\mathrm{#1})}{=}}
\newcommand{\oleq}[1]{\overset{(\mathrm{#1})}{\leq}}
\newcommand{\ogeq}[1]{\overset{(\mathrm{#1})}{\geq}}

\usepackage{cite}
\usepackage{printlen}
\usepackage{verbatim}

\DeclareMathOperator{\supp}{supp}

\begin{document}
\maketitle

\begin{abstract}
In this work, arithmetic distribution matching (ADM) is presented. ADM invertibly transforms a discrete memoryless source (DMS) into a target DMS. ADM can be used for probabilistic shaping and for rate adaption. Opposed to existing algorithms for distribution matching, ADM works online and can transform arbitrarily long input sequences. It is shown analytically that as the input length tends to infinity, the ADM output perfectly emulates the target DMS with respect to the normalized informational divergence and the entropy rate. Numerical results are presented that confirm the analytical bounds.
\end{abstract}

\section{Introduction}

\emph{Distribution matching} transforms the output of a discrete memoryless source (DMS) into a sequence that emulates a target DMS, see Figure \ref{fig: DMS} for an illustration. The transformation of a distribution matcher is invertible, i.e., the input can be recovered from the output. Distribution matching is used for example for probabilistic shaping \cite[Section~IV.A]{forney1984efficient},\cite{bocherer2012capacity} and for rate adaption \cite[Section~VI.]{mackay1999good}.

Distribution matchers can be implemented using variable length coding. In \cite{bocherer2011matching} and \cite{amjad2013fixed}, algorithms for optimal variable-to-fixed (v2f) length and fixed-to-variable (f2v) length matching are presented, respectively. The drawback of these optimal matchers is that the complete codebook needs to be calculated offline, which is infeasible for large codebook sizes. For data compression, Huffman \cite{huffman1952method} and Tunstall \cite{tunstall1967synthesis} codes have a similar problem. Arithmetic codes for data compression \cite{rissanen1979arithmetic,witten1987arithmetic} are sub-optimal variable length codes where encoding and decoding can be done online, i.e., no codebook needs to be stored. The use of arithmetic coding for distribution matching was proposed in \cite[Appendix G]{mackay1999good}. However, as stated by the authors of \cite{mackay1999good}, their algorithm is incomplete, in particular, it is not invertible in the provided description. The authors in \cite{han1997interval} propose a non-invertible algorithm for exact random number generation based on the idea of arithmetic coding.   

The main contributions of this work are the development and the analysis of an algorithm for arithmetic distribution matching (ADM). We review f2v length distribution matching in Section~\ref{sec:preliminaries}. In particular, we discuss in Section~\ref{subsec:compression} its relation to data compression. We then present in Section~\ref{sec:arithmetic} our ADM algorithm. We theoretically analyze the performance of our algorithm in Section~\ref{sec:analysis}. In particular, we show that as the input length tends to infinity, the output of ADM perfectly emulates the target DMS with respect to normalized informational divergence and entropy rate. We provide numerical results in Section~\ref{sec:numerical} that confirm our analytical bounds. Our implementation is available at \cite{website:ADM} and was used in \cite{bocherer2014probabilistic} for coded modulation with probabilistic shaping.

\section{Fixed-to-variable length Distribution Matching}
\label{sec:preliminaries}

\begin{figure}
\centering
\begin{tikzpicture}[scale = 0.5]

\draw (0,0) -- (3,0);
\draw (3,0) -- (3,2);
\draw (3,2) -- (0,2);
\draw (0,2) -- (0,0);
\draw [->] (3,1) -- (6,1);
\draw (6,0) -- (9,0);
\draw (9,0) -- (9,2);
\draw (9,2) -- (6,2);
\draw (6,2) -- (6,0);
\draw [->] (9,1) -- (12,1);
\node at (1.5,1) {DMS $P_S$};
\node at (7.5,1) {matcher};
\node at (4.5,1.5) {$s_1,s_2,\ldots$};
\node at (11,1.5) {$c_1,c_2,\ldots$};

\draw[dashed] (-0.5,-0.5) -- (9.5,-0.5);
\draw[dashed] (-0.5,2.5) -- (9.5,2.5);
\draw[dashed] (-0.5,-0.5) -- (-0.5,2.5);
\draw[dashed] (9.5,-0.5) -- (9.5,2.5);

\draw (0,-3) -- (3,-3);
\draw (3,-3) -- (3,-1);
\draw (3,-1) -- (0,-1);
\draw (0,-1) -- (0,-3);
\draw [->] (3,-2) -- (6,-2);
\node at (1.5,-2) {DMS $P_Z$};
\node at (4.5,-1.5) {$z_1,z_2,\ldots$};

\end{tikzpicture}
\caption{The output $s_1,s_2,\dotsc$ of the DMS $P_S$ is transformed by the distribution matcher. The output sequence $c_1,c_2,\dotsc$ appears similar to the output sequence $z_1,z_2,\dotsc$ of the target DMS $P_Z$. As indicated by the dashed box, the source $P_S$ together with the matcher emulates the target DMS $P_Z$.}
\label{fig: DMS}
\end{figure}
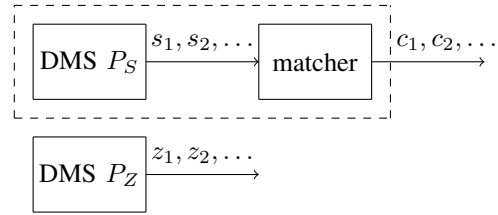

The concept of distribution matching is illustrated in Figure~\ref{fig: DMS}. We describe in the following f2v matching. For clarity of exposure, we consider binary input and binary output. We denote random variables by capital letters $S$ and realizations by small letters $s$. A binary DMS $P_S$ generates a bit sequence $\boldsymbol{S}=S_1,S_2,\ldots,S_n$ of fixed length $n$. The bits $S_i$ are \emph{independent and identically distributed} (iid) according to the distribution $P_S(0) = p_\mathrm{src}$ and $P_S(1) = 1-p_\mathrm{src}$. Suppose the source output is $\boldsymbol{s}$. The matcher transforms $\boldsymbol{s}$ into a binary sequence $\boldsymbol{c}=c_1,c_2,\ldots,c_{\ell(\boldsymbol{c})}$ of variable length $\ell(\boldsymbol{c})$, i.e., the matcher outputs codewords of a f2v length codebook $\setc$. The goal of distribution matching is to emulate a binary DMS with an arbitrary but fixed target distribution $P_Z(0) = p_\mathrm{code}$ and $P_Z(1) = 1-p_\mathrm{code}$. We explain in the next paragraphs what we mean by ``emulation''.

\subsection{Interval representation}
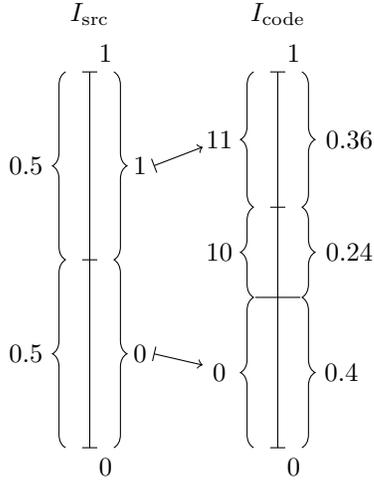
\begin{figure}
\centering
\begin{tikzpicture}[scale = 0.5]

\node [anchor = south] at (1,11) {$I_\mathrm{src}$};
\draw (1,0) -- (1,10);

\draw (0.8,0) -- (1.2,0);
\node [anchor = west] at (1,-0.5) {$0$};
\draw (0.8,10) -- (1.2,10);
\node [anchor = west] at (1,10.5) {$1$};
\draw (0.8,5)  -- (1.2,5);
%\node [anchor = west] at (1.2,5) {$p_\mathrm{src}$};
%\node [anchor = east] at (0,2.5) {$0$};
%\node [anchor = east] at (0,7.5) {$1$};

\draw [decorate,decoration={brace,amplitude=5pt},xshift=-4pt,yshift=0pt]
(0.5,0) -- (0.5,5)node [black,midway,xshift = -15pt] {$0.5$};
\draw [decorate,decoration={brace,mirror,amplitude=5pt},xshift=4pt,yshift=0pt]
(1.5,0) -- (1.5,5)node [black,midway,xshift = 10pt] {$0$};

\draw [decorate,decoration={brace,amplitude=5pt},xshift=-4pt,yshift=0pt]
(0.5,5) -- (0.5,10)node [black,midway,xshift = -15pt] {$0.5$};
\draw [decorate,decoration={brace,mirror,amplitude=5pt},xshift=4pt,yshift=0pt]
(1.5,5) -- (1.5,10)node [black,midway,xshift = 10pt] {$1$};

\draw [decorate,decoration={brace,amplitude=5pt},xshift=-4pt,yshift=0pt]
(5.5,0) -- (5.5,4)node [black,midway,xshift = -13pt] {$0$};
\draw [decorate,decoration={brace,mirror,amplitude=5pt},xshift=4pt,yshift=0pt]
(6.5,0) -- (6.5,4)node [black,midway,xshift = 15pt] {$0.4$};

\draw [decorate,decoration={brace,amplitude=5pt},xshift=-4pt,yshift=0pt]
(5.5,4) -- (5.5,6.4)node [black,midway,xshift = -13pt] {$10$};
\draw [decorate,decoration={brace,mirror,amplitude=5pt},xshift=4pt,yshift=0pt]
(6.5,4) -- (6.5,6.4)node [black,midway,xshift = 18pt] {$0.24$};

\draw [decorate,decoration={brace,amplitude=5pt},xshift=-4pt,yshift=0pt]
(5.5,6.4) -- (5.5,10)node [black,midway,xshift = -13pt] {$11$};
\draw [decorate,decoration={brace,mirror,amplitude=5pt},xshift=4pt,yshift=0pt]
(6.5,6.4) -- (6.5,10)node [black,midway,xshift = 18pt] {$0.36$};

\node [anchor = south] at (6,11) {$I_\mathrm{code}$};
\draw (6,0) -- (6,10);

\draw (5.8,0) -- (6.2,0);
\node [anchor = west] at (6,-0.5) {$0$};
\draw (5.8,10) -- (6.2,10);
\node [anchor = west] at (6,10.5) {$1$};
\draw (5.4,4) -- (6.6,4);
%\node [anchor = west] at (5.8,4) {$p_\mathrm{code}$};
%\draw (4.8,1.6) -- (5.2,1.6);
%\node [anchor = west] at (5.2,1.6) {$p_\mathrm{code}^{2}$};
\draw (5.8,6.4) -- (6.2,6.4);
%\node [anchor = west] at (5.2,6.4) {$p_\mathrm{code}+p_\mathrm{code}\cdot(1-p_\mathrm{code})$};
%\node [anchor = east] at (4.3,1) {$00$};
%\node [anchor = east] at (4.3,3) {$01$};
%\node [anchor = east] at (4.3,5) {$10$};
%\node [anchor = east] at (4.3,8) {$11$};

\draw[|->] (2.7,7.5) -- (4,8);
\draw[|->] (2.7,2.5) -- (4,2.2);

\end{tikzpicture}
\caption{The source is $P_S(0)=P_S(1)=0.5$ and the target DMS is defined by $P_Z(0)=1-P_Z(1)=0.4$. The matcher maps $1$ to $11$ and $0$ to $0$. The sequences $11$, $10$ and $0$ appear at the matcher output with probability $0.5$, $0$, and $0.5$, respectively. At the output of the target DMS, $11$, $10$, and $0$ would appear with probability $0.36$, $0.24$, and $0.4$, respectively.}
\label{fig: prel}
\end{figure}
We represent the probabilities of the input realizations $\boldsymbol{s}$ and the target probabilities of the output realizations $\boldsymbol{c}$ by subintervals of the interval $[0;1)$. We denote the subinterval representing the probability of $\textbf{s}$ by $I_\mathrm{src}$ and the subinterval representing the target probability of $\vecc$ by $I_\mathrm{code}$. We identify the matcher input $\boldsymbol{s}$ with $I_\mathrm{src}$ and the matcher output $\boldsymbol{c}$ with $I_\mathrm{code}$, i.e., the matcher maps $I_\mathrm{src}$ to $I_\mathrm{code}$. We display an example of the interval representation in Figure \ref{fig: prel}.

\subsection{Informational Divergence}
The division of the interval $[0;1)$ into subintervals defines the variable length codebook $\setc$. For the example in Figure~\ref{fig: prel}, the codebook is $\setc=\{11,10,0\}$. For the input length $n$, the matcher uses $2^n$ output bitsequences with non-zero probability. In Figure~\ref{fig: prel}, $n=1$ and the two possible output sequences are $\{11,0\}$. The mapping performed by the matcher defines a probability distribution on $\cal{C}$. We represent the matcher output taking values in $\setc$ by the random variable $Y$. The codeword $\boldsymbol{c}$ appears at the matcher output with probability
\begin{align}
P_Y(\boldsymbol{c}) = \begin{cases} P^{n}_{X}(\boldsymbol{s}), & \text{if input $\boldsymbol{s}$ maps to $\boldsymbol{c}$}\nonumber\\ 0, & \text{if no $\boldsymbol{s}$ maps to $\boldsymbol{c}$.} \end{cases}
\end{align}
The target DMS would have put out the codeword $\boldsymbol{c}$ with probability
\begin{align}
P_Z^{\cal{C}}(\boldsymbol{c}) = \prod \limits_{i = 1}^{\ell(\boldsymbol{c})} P_Z(c_i).\nonumber
\end{align}
We say that $P_Z$ induces the distribution $P_Z^{\setc}$ on the codebook $\setc$. In our interval representation, the probability $P_Z^{\setc}(\boldsymbol{c})$ by which the target DMS $P_Z$ would have generated $\boldsymbol{c}$ is represented by the interval size of $I_\mathrm{code}$ and the actual probability is represented by the interval size of $I_\mathrm{src}$. The matcher output is a good approximation of the target DMS output if $P_Y(\boldsymbol{c})\approx P_Z^{\setc}(\boldsymbol{c})$ and equivalently if $I_\mathrm{src}$ and $I_\mathrm{code}$ have approximately the same size. This intuition is formalized by the \emph{informational divergence} of $P_Y$ and $P_Z^{\setc}$, which is defined by
\begin{align}
\kl(P_Y\Vert P_Z^{\cal{C}})&=\sum_{\vecc\in\supp P_Y}P_Y(\vecc)\log_2\frac{P_Y(\vecc)}{P_Z^{\cal{C}}(\vecc)}\nonumber\\
&=\sum_{\vecc\in\supp P_Y}P_Y(\vecc)\log_2\frac{I_\mathrm{src}(\vecc)}{I_\mathrm{code}(\vecc)}
\label{eq: divergence}
\end{align}
where $\supp P_Y=\{\vecc\in\setc\colon P_Y(\vecc)>0\}$ denotes the support of $P_Y$. We can see that the informational divergence depends on the ratio $I_\mathrm{src}/I_\mathrm{code}$. It is small if $I_\mathrm{src}$ and $I_\mathrm{code}$ have approximately the same size for all codewords that occur with non-zero probability.

\subsection{Optimal distribution matching}

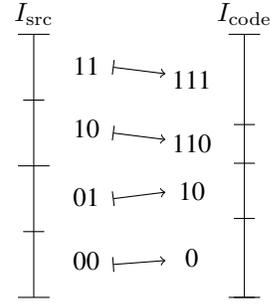
\begin{figure}
\centering
\begin{tikzpicture}[scale = 0.7]

\draw (0,0) -- (0,5);
\draw (4,0) -- (4,5);
\draw(-0.3,0) -- (0.3,0);
\draw(-0.2,1.25) -- (0.2,1.25);
\draw(-0.3,2.5) -- (0.3,2.5);
\draw(-0.2,3.75) -- (0.2,3.75);
\draw(-0.3,5) -- (0.3,5);

\node at (1,0.7) {00};
\node at (1,1.9) {01};
\node at (1,3.2) {10};
\node at (1,4.4) {11};

\node[anchor = south] at (0,5) {$I_\mathrm{src}$};
\node[anchor = south] at (4,5) {$I_\mathrm{code}$};

\draw(3.7,0) -- (4.3,0);
\draw(3.7,5) -- (4.3,5);

\draw (3.8,0) -- (4.2,0);
\draw (3.8,1.5) -- (4.2,1.5);
\node at (3,0.75) {0};
\draw (3.8,2.55) -- (4.2,2.55);
\node at (3,2.025) {10};
\draw (3.8,3.285) -- (4.2,3.285);
\node at (3,2.9175) {110};
\node at (3,4.1425) {111};

\draw[|->] (1.5,1.25/2) -- (2.5,1.4/2);
\draw[|->] (1.5,3.75/2) -- (2.5,4/2);
\draw[|->] (1.5,6.25/2) -- (2.5,6.0/2);
\draw[|->] (1.5,8.75/2) -- (2.5,8.5/2);

\end{tikzpicture}
\caption{An optimal code for input length $n = 2$, $P_S(0) = 0.5$ and $P_Z(0) = 0.3$. The resulting informational divergence is $\kl(P_Y\Vert P_Z^{\cal{C}})=0.074584$.}
\label{fig: optimal f2v int}
\end{figure}

In \cite{amjad2013fixed} an algorithm is described to generate f2v length codes for distribution matching that minimize \eqref{eq: divergence}. In Figure~\ref{fig: optimal f2v int} we show an example for such an optimal code. The input length is $n=2$, the source is uniform and the target DMS is $P_Z(0)=1-P_Z(1)=0.3$ The codebook $\cal{C}$ consists of all 4 subintervals of $[0,1)$ displayed in Figure~\ref{fig: optimal f2v int}.
%, so this code can be represented by a complete tree, where a complete tree is a tree without unused leaves. Figure \ref{fig: complete tree} shows this tree.
The resulting informational divergence is 0.0746. The optimal mapping has to be calculated offline and stored. The required memory increases exponentially with the input length and becomes impractical already for reasonably small input lengths.
%Furthermore, the algorithm presented in \cite{amjad2013fixed} only works for uniform input distributions. 

%\tikzstyle{transition}=[circle,draw=black!50,fill=black!20,thick,inner sep=0pt,minimum size=3mm]
%\tikzstyle{leaf}=[circle,draw=green!50!black,fill=green!20,thick,inner sep=0pt,minimum size=4mm]
%\tikzstyle{emptyleaf}=[circle,draw=red!50,fill=red!20,thick,inner sep=0pt,minimum size=2mm]
%\begin{figure}[ht]
%\begin{center}
%\begin{tikzpicture}[scale=0.5,->,>=stealth',level 1/.style={sibling distance=3cm,level distance=1.5cm},
%    level 2/.style={sibling distance=2.2cm, level distance=1.5cm},
%    level 3/.style={sibling distance=1.1cm, level distance=1.5cm}]
%\node[transition] {}
%    child {node[leaf] {00}}
%    child {node[transition] {}
%        child{node[leaf]{01}}
%        child{node[transition] {}
%	    child{node[leaf] {10}}
%	    child{node[leaf] {11}}
%	}	
%    };

%\node[anchor = south west] at (1,-1) {1};
%\node[anchor = south east] at (-1,-1) {0};
%\end{tikzpicture}
%\end{center}
%\caption{The optimal code is represented by a complete tree.}
%\label{fig: complete tree}
%\end{figure} 

\subsection{Preview: Arithmetic distribution matching}
\begin{figure}
\centering
\begin{tikzpicture}[scale = 0.7]

\draw (0,0) -- (0,5);
\draw (4,0) -- (4,5);
\draw(-0.3,0) -- (0.3,0);
\draw(-0.2,1.25) -- (0.2,1.25);
\draw(-0.3,2.5) -- (0.3,2.5);
\draw(-0.2,3.75) -- (0.2,3.75);
\draw(-0.3,5) -- (0.3,5);

\node at (1,0.7) {00};
\node at (1,1.9) {01};
\node at (1,3.2) {10};
\node at (1,4.4) {11};

\node[anchor = south] at (0,5) {$I_\mathrm{src}$};
\node[anchor = south] at (4,5) {$I_\mathrm{code}$};

\draw(3.7,0) -- (4.3,0);
\draw(3.7,5) -- (4.3,5);

\draw (3.8,0.135) -- (4.2,0.135);
\draw (3.8,0.45) -- (4.2,0.45);
\node at (3,0.3) {001};
\draw (3.8,1.5) -- (4.2,1.5);
\draw (3.8,1.815) -- (4.2,1.815);
\node at (3,1.65) {100};
\draw (3.8,2.55) -- (4.2,2.55);
\draw (3.8,3.285) -- (4.2,3.285);
\node at (3,2.9) {110};
\draw (3.8,3.7995) -- (4.2,3.7995);
\draw (3.8,4.1597) -- (4.2,4.1597);
\node at (3,3.97) {11110};

\draw[|->] (1.5,1.25/2) -- (2.5,0.7/2);
\draw[|->] (1.5,3.75/2) -- (2.5,3.3/2);
\draw[|->] (1.5,6.25/2) -- (2.5,6.0/2);
\draw[|->] (1.5,8.6/2) -- (2.2,8.2/2);

\end{tikzpicture}
\caption{Example of a code generated by an arithmetic matcher where $n = 2$, $P_S(0)=0.5$ and $P_Z(0) = 0.3$. The resulting informational divergence is $\kl(P_Y\Vert P_Z^{\cal{C}})=1.6346$, which is larger than the divergence of the optimal code in Figure~\ref{fig: optimal f2v int}.}
\label{fig: codeint}
\end{figure}
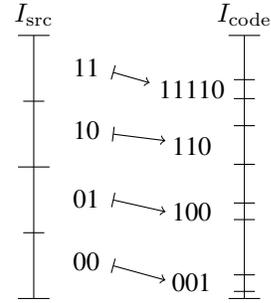
The main idea of ADM is to require that the code interval identifies the source interval, i.e., $I_\mathrm{code} \subseteq I_\mathrm{src}$. By this requirement, we also give up on using all subintervals of $[0,1)$ as codewords. For the same example as in Figure \ref{fig: optimal f2v int}, we display in Figure \ref{fig: codeint} a matcher with the $I_\mathrm{code} \subseteq I_\mathrm{src}$ property. The informational divergence of the matcher in Figure~\ref{fig: codeint} is equal to $1.6346$, which is larger than the divergence of the optimal code in Figure~\ref{fig: optimal f2v int}, so the code with the $I_\mathrm{code} \subseteq I_\mathrm{src}$ property performs worse than the optimal code. However, as we will see in the remaining sections, ADM allows us to encode and decode online for arbitrarily long input sequences. Furthermore, we will see that for long input sequences, ADM results in a smaller divergence than repeatedly applying an optimal code.

\subsection{Compression decoder as a matching encoder}\label{subsec:compression}
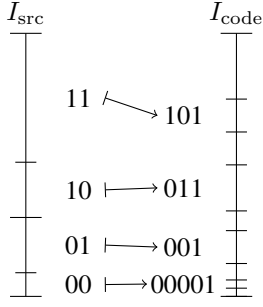
\begin{figure}
\centering
\begin{tikzpicture}[scale = 0.7]

\draw (0,0) -- (0,5);
\draw (4,0) -- (4,5);
\draw(-0.3,0) -- (0.3,0);
\draw(-0.2,0.45) -- (0.2,0.45);
\draw(-0.3,1.5) -- (0.3,1.5);
\draw(-0.2,2.55) -- (0.2,2.55);
\draw(-0.3,5) -- (0.3,5);

\node at (1,0.225) {00};
\node at (1,0.975) {01};
\node at (1,2.025) {10};
\node at (1,3.775) {11};

\node[anchor = south] at (0,5) {$I_\mathrm{src}$};
\node[anchor = south] at (4,5) {$I_\mathrm{code}$};

\draw(3.7,0) -- (4.3,0);
\draw(3.7,5) -- (4.3,5);

\draw (3.8,0) -- (4.2,0);
\draw (3.8,0.15625) -- (4.2,0.15625);
\draw (3.8,0.3125) -- (4.2,0.3125);
\node at (3,0.23438) {00001};
\draw (3.8,0.625) -- (4.2,0.625);
\draw (3.8,1.25) -- (4.2,1.25);
\node at (3,0.93750) {001};
\draw (3.8,1.625) -- (4.2,1.625);
\draw (3.8,2.5) -- (4.2,2.5);
\node at (3,2.0625) {011};
\draw (3.8,3.125) -- (4.2,3.125);
\draw (3.8,3.75) -- (4.2,3.75);
\node at (3,3.4375) {101};

\draw[|->] (1.5,0.225) -- (2.3,0.23438);
\draw[|->] (1.5,0.975) -- (2.5,0.9375);
\draw[|->] (1.5,2.025) -- (2.5,2.0625);
\draw[|->] (1.5,3.775) -- (2.5,3.4375);

\end{tikzpicture}
\caption{Example of an arithmetic source compression code for input length $n = 2$, $P_S(0) = 0.3$ and $P_Z(0) = 0.5$.}
\label{fig: srccoding}
\end{figure}
It is claimed in \cite{mackay1999good} that an ADM for a target DMS $P_Z$ can be realized by applying the decoder of an arithmetic source compression code for $P_Z$ to the output of a uniform source $P_S$. We illustrate that this is not possible by an example. We consider input length $n=2$ and a source $P_S(0)=1-P_S(1)=0.3$. We compress $P_S$ by emulating the uniform target DMS $P_Z(0)=P_Z(1)=0.5$. The resulting arithmetic source compression code is displayed in Figure~\ref{fig: srccoding}. Suppose now we want to apply the corresponding decoder to the output of a uniform source. This means we apply the inverse mapping from right to left. The output $101$ maps to $11$, so this is fine. However, if the output is 11, this approach fails, since the encoder maps nothing to 11, so 11 cannot be encoded. We conclude that in general, an arithmetic decoder cannot be used as an encoder.

\section{Arithmetic Distribution Matching}
\label{sec:arithmetic}
We describe the algorithms for an ADM encoder and decoder for binary input distributions.

\subsection{Basic operations}
There are two basic operations, which are used by the encoder as well as the decoder.

\subsubsection{Read Bits}
\label{sec: read}
A bitsequence that was created by a DMS can be represented by an interval by successively reading its bits.
We start with the interval $I = [0;1)$. Now we divide the interval $I$ in two parts according to the probability distribution $p$ of the corresponding DMS. The lower subinterval $[0;p)$ is assigned to $0$, the upper subinterval $[p;1)$ is assigned to $1$. Then we read the first bit of the bitsequence. If there is a $0$, we choose the lower subinterval as the new interval $I$, if there is a $1$, we choose the upper subinterval. This interval $I$ represents the bitsequence we have read.
We continue this process recursively by subdividing the current interval $I$ according to $p$. When all bits of the bitsequence are read, $I$ represents the whole bitsequence. 

\subsubsection{Refine Candidate List}
\label{sec: refine}
We subdivide the interval $[0;1)$ according to $p$. In this way we create two subintervals which we call candidates. The lower subinterval $[0;p)$ is assigned to $0$, the upper subinterval $[p;1)$ is assigned to $1$.
%By continuing the subdivision of both candidates, we create a list of candidates representing bitsequences. 
We call this process \emph{refinement}. All intervals created by this process can also be refined by subdividing them equivalently according to $p$. So after a second refinement we have four candidates.
%In each refinement step, the new bits are concatenated with the preceding bitsequences.

\subsection{Encoder}

\label{sec: encoder}
\begin{comment}
At first the encoder carries out one refinement of $I_\mathrm{code}$ using $p_\mathrm{code}$ and so creates a candidate list as described in \ref{sec: refine}. Then the encoder reads a bit from $s$ and adjusts $I_\mathrm{src}$ using $p_\mathrm{src}$ as described in \ref{sec: read}. The encoder checks whether one of the intervals of the candidate list contains $I_\mathrm{src}$. If no candidate contains $I_\mathrm{src}$, the encoder reads the next bit and adjusts $I_\mathrm{src}$ until a candidate contains $I_\mathrm{src}$. While there is a candidate which contains $I_\mathrm{src}$, the bitsequence corresponding to the candidate is written in the output buffer of $c$. Additionally the encoder assigns the candidate which contains $I_\mathrm{src}$ to $I_\mathrm{code}$ and carries out one refinement of $I_\mathrm{code}$ thus creating a new candidate list. These steps are repeated until there is no candidate, which contains $I_\mathrm{src}$. Then the encoder reads the next bit from $s$ and repeats the process, until all $n$ bits of $s$ are read.
After this procedure  $I_\mathrm{code}$ contains $I_\mathrm{src}$ and so $I_\mathrm{src}$ identifies $I_\mathrm{code}$. In order to decode $c$, it is necessary, that $I_\mathrm{src}$ contains $I_\mathrm{code}$. In this way $c$ identifies $s$. The encoder refines the candidate list for $I_\mathrm{code}$ until $I_\mathrm{src}$ contains one of the candidates. Then the encoder writes the  bitsequence corresponding to this candidate in the output buffer of $c$.
\end{comment}

\begin{algorithm}
\caption{Encoder}\label{enc}
\begin{algorithmic}[1]
\State $\vecs \gets \text{input sequence}$
\State $p_\mathrm{src} \gets \text{source distribution}$
\State $p_\mathrm{code} \gets \text{target distribution}$
\State $ \text{candidateList} = \text{refine}([0;1),p_\mathrm{code})$
\State $\vecc = \text{empty array}$
\For{$i$=1 \textbf{to} \text{length($\vecs$)}}
\State $I_\mathrm{src} = \text{readBit}(I_\mathrm{src},p_\mathrm{src},s_i)$
\While{$\exists j \colon I_\mathrm{src} \subseteq \text{candidateList}(j)$}
\State \text{append bits corresponding to $j$ to $\vecc$}
\State $I_\mathrm{code} = \text{refine}(\text{candidateList}(j),p_\mathrm{code})$
\EndWhile
\EndFor
\State $[I_\mathrm{code}^u,I_\mathrm{code}^l] = \text{refine}(I_\mathrm{code},p_\mathrm{code})$
\While{$\nexists k \colon I_\mathrm{code}^u(k)  \subseteq I_\mathrm{src}$}
\State $I_\mathrm{code}^u = \text{refine}(I_\mathrm{code}^u,p_\mathrm{code})$
\EndWhile
\While{$\nexists l \colon I_\mathrm{code}^l(l)  \subseteq I_\mathrm{src}$}
\State $I_\mathrm{code}^l = \text{refine}(I_\mathrm{code}^l,p_\mathrm{code})$
\EndWhile
\State $I_\mathrm{code} = \text{max}(I_\mathrm{code}^u(k),I_\mathrm{code}^l(l))$
\State $\text{append corresponding bits to $\vecc$}$
\end{algorithmic}
\end{algorithm}
The arithmetic encoder creates a candidate list by refining the interval $I_\mathrm{code}$ using $p_\mathrm{code}$. Then it reads bits of $\textbf{s}$ until the corresponding interval $I_\mathrm{src}$ identifies one of the candidates. This candidate is the new $I_\mathrm{code}$ and $I_\mathrm{src} \subseteq I_\mathrm{code}$. The bit corresponding to this candidate is written in the output buffer of $\vecc$. Then the encoder refines $I_\mathrm{code}$. If $I_\mathrm{src}$ identifies one of the new candidates, the corresponding bit is written in the output buffer of $\vecc$ too and this candidate is the new $I_\mathrm{code}$ again. The encoder continues the refinement of $I_\mathrm{code}$ and puts out the corresponding bits if possible, until $I_\mathrm{src}$ is not contained in any of the new candidates. Then the encoder starts over again by reading the next bit. It repeats this process until all bits are read.

To finish the encoding, when all input bits are read there is an upper candidate $I_\mathrm{code}^u$ and a lower candidate $I_\mathrm{code}^l$. Then the encoder refines $I_\mathrm{code}^u$ until one candidate identifies $I_\mathrm{src}$. It then refines $I_\mathrm{code}^l$ until a second candidate identifies $I_\mathrm{src}$. Then the encoder chooses the larger one of the two candidates as the new $I_\mathrm{code}$. So $I_\mathrm{code} \subseteq I_\mathrm{src}$ holds. Finally it appends the additional bits corresponding to this candidate to $\vecc$. This finalization is necessary to guarantee decodability.

Algorithm~\ref{enc} shows a pseudocode for the encoder.

\subsection{Decoder}

\begin{algorithm}
\caption{Decoder}\label{dec}
\begin{algorithmic}[1]
\State $\vecc \gets \text{codeword}$
\State $p_\mathrm{src} \gets \text{source distribution}$
\State $p_\mathrm{code} \gets \text{target distribution}$
\State $n \gets \text{length of input sequence $\vecs$}$
\State $ \text{candidateList} = \text{refine}([0;1),p_\mathrm{src})$
\State $\vecs = \text{empty string}$
\State $i = 1$
\While{$\text{length}(\vecs) < n$}
\State $I_\mathrm{code} = \text{readBit}(I_\mathrm{code},p_\mathrm{code},c_i)$
\State $i = i + 1$
\While{$\exists j \colon I_\mathrm{code} \subseteq \text{candidateList}(j)$}
\State \text{append bits corresponding to $j$ to $\vecs$}
\State $I_\mathrm{src} = \text{refine}(\text{candidateList}(j),p_\mathrm{src})$
\EndWhile
\EndWhile
\end{algorithmic}
\end{algorithm}

\label{sec: decoder}
\begin{comment}
The decoder carries out one refinement of $I_\mathrm{src}$ using $p_\mathrm{src}$ and so creates a candidate list. It then reads a bit from $c$ and adjusts $I_\mathrm{code}$ using $p_\mathrm{code}$ as described in \ref{sec: read}. The decoder checks whether one of the intervals of the candidate list contains $I_\mathrm{code}$. If no candidate contains $I_\mathrm{code}$, the decoder reads the next bit and adjusts $I_\mathrm{code}$ until a candidate contains $I_\mathrm{code}$. While there is a candidate which contains $I_\mathrm{code}$, the bitsequence corresponding to the candidate is written in the output buffer of $s$. Additionally the decoder assigns the candidate which contains $I_\mathrm{code}$ to $I_\mathrm{src}$ and carries out one refinement of $I_\mathrm{src}$ thus creating a new candidate list. These steps are repeated until there is no candidate, which contains $I_\mathrm{code}$. Then the decoder reads the next bit from $c$ and repeats the process, until all $n$ bits of $s$ are written in its output buffer.
\end{comment}

The arithmetic decoder creates a candidate list by refining the interval $I_\mathrm{src}$ using $p_\mathrm{src}$. It then reads bits of $\vecc$ until $I_\mathrm{code}$ identifies one of the candidates. The bit corresponding to this candidate is written in the output buffer of $\textbf{s}$. This candidate is the new $I_\mathrm{src}$. It is refined, and if $I_\mathrm{code}$ identifies one of the new candidates, the corresponding bit is written in the output buffer of $\textbf{s}$ too. Again, this candidate is the new $I_\mathrm{src}$. This is repeated until $I_\mathrm{code}$ does not identify any of the candidates. Then the decoder starts reading bits from $\vecc$ again. It carries on until $n$ bits are written to the output buffer of $\textbf{s}$.

Algorithm~\ref{enc} shows a pseudocode for the decoder.
\begin{comment}
The arithmetic decoder does the same as the encoder, but the roles of $\textbf{s}$ and $\textbf{c}$, $I_\mathrm{src}$ and $I_\mathrm{code}$ as well as the roles of $p_\mathrm{src}$ and $p_\mathrm{code}$ are exchanged. The stopping criterion is satisfied when all $n$ bits of $\textbf{s}$ are written in the output buffer.
\end{comment}

\subsection{Implementation}
For the floating point implementation of the algorithms described above we have to prevent the subintervals from becoming too small for a representation in floating point numbers. That is why they are repeatedly scaled during the encoding and decoding process. Additionally the decoder has to execute the same scalings as the encoder, to avoid different rounding at the encoder and the decoder, i.e. the decoder needs to emulate the encoder exactly in terms of floating point operations.
\begin{comment}
\subsection{Variable to Fixed length Arithmetic Coding}
We can modify the fixed to variable length encoder and decoder described above to a variable to fixed length encoder and decoder. The encoder works in the same way as described, but the termination condition is different. The algorithm stops reading an additional bit, if by reading this bit, the two possible subintervals could not be identified by subintervals of $I_\mathrm{code}$ corresponding to sequences of the fixed output length m. Thus the encoder creates a uniquely decodable code.
\end{comment}

\section{Analysis}
\label{sec:analysis}
\subsection{Informational Divergence}
Suppose the output of the encoder is the sequence $\vecc$. The width of the source interval is equal to the probability that $\vecc$ is generated, i.e., $I_\mathrm{src}(\vecc)=P_Y(\vecc)$. The width of the code interval $I_\mathrm{code}$ is equal to the probability by which the target DMS would generate $\vecc$, i.e., $I_\mathrm{code}(\vecc)=P_Z^{\cal{C}}(\vecc)$. 
\begin{proposition}\label{prop:ratiobound}
The ratio of the interval sizes is bounded as
\begin{align}
1\leq\frac{I_\mathrm{src}(\vecc)}{I_\mathrm{code}(\vecc)}=\frac{P_Y(\vecc)}{P_Z^{\cal{C}}(\vecc)}\leq\frac{1}{p_\mathrm{code}\cdot (1-p_\mathrm{code})}.\nonumber
\end{align}
\end{proposition}
\begin{IEEEproof}
The left inequality holds since the algorithm guarantees $I_\mathrm{code} \subseteq I_\mathrm{src}$. The upper bound is proved in the appendix.
\end{IEEEproof}
We bound the informational divergence
\begin{align}
\kl(P_Y\Vert P_Z^{\cal{C}})&=\sum_{\vecc\in \supp P_Y}P_Y(\vecc)\log_2\frac{P_Y(\vecc)}{P_Z^{\cal{C}}(\vecc)}\nonumber\\
&\leq\sum_{\vecc\in \supp P_Y}P_Y(\vecc)\log_2\frac{1}{p_\mathrm{code}\cdot (1-p_\mathrm{code})}\nonumber\\
&=\log_2\frac{1}{p_\mathrm{code}\cdot (1-p_\mathrm{code})}\label{eq:klbound}
\end{align}
Thus the un-normalized informational divergence is bounded from above by a constant that does not depend on the input length $n$. The expected output length is bounded as
\begin{align}
\expop[\ell(Y)] \ogeq{a} \entop(P_{Y}) \oeq{b} \entop(P_{S^n}) = n\entop(P_S)\label{eq:lengthbound}
\end{align}
where (a) follows by the converse of the source coding theorem \cite[Theorem~5.3.1]{cover2006elements} and where (b) holds because the mapping of the matcher is one-to-one. We can now bound the normalized informational divergence as
\begin{align}
\frac{\kl(P_Y\Vert P_Z^{\cal{C}})}{\expop[\ell(Y)]}&\oleq{a} \frac{\log_2\frac{1}{p_\mathrm{code}\cdot (1-p_\mathrm{code})}}{\expop[\ell(Y)]}\nonumber\\
&\oleq{b}\frac{\log_2\frac{1}{p_\mathrm{code}\cdot (1-p_\mathrm{code})}}{n\entop(P_{S})}\label{eq: normalized}
\end{align}
where (a) follows by Proposition~\ref{prop:ratiobound} and where (b) follows by \eqref{eq:lengthbound}. We can see from \eqref{eq: normalized} that the normalized informational divergence approaches zero for large input lengths $n$. Our numerical results in Section~\ref{sec:numerical} confirm this observation.

\subsection{Rate}

\begin{proposition}\label{prop:rate}
As the input length $n$ tends to infinity, the entropy rate of the matcher output converges to the entropy of the target distribution, i.e.,
\begin{align}
\label{rate to entropy}
\left| \frac{\entop(P_Y)}{\expop[\ell(Y)]} - \entop(P_Z) \right| \overset{n\to\infty}{\rightarrow} 0.
\end{align}
\end{proposition}
\begin{IEEEproof}
According to \eqref{eq: normalized} 
\begin{align}
\frac{\kl(P_Y\Vert P_Z^{\cal{C}})}{\expop[\ell(Y)]} \overset{n\to\infty}{\rightarrow} 0 \nonumber
\end{align}
which according to \cite[Proposition 6] {bocherer2013informational} implies \eqref{rate to entropy}.
\end{IEEEproof}
Since the mapping is one-to-one, we have $\entop(P_Y)=n\entop(P_S)$. In the average, $n$ input bits are transformed into $\expop[\ell(Y)]$ output bits. In terms of the conversion rate $n/\expop[\ell(Y)]$, Proposition~\ref{prop:rate} states that
\begin{align}
\left| \frac{n}{\expop[\ell(Y)]} - \frac{\entop(P_Z)}{\entop(P_S)} \right| \overset{n\to\infty}{\rightarrow} 0. \nonumber
\end{align}

\subsection{Binary data compression} 

We now want to show how ADM can be used for data compression. Suppose $P_Z(0)=P_Z(1)=\frac{1}{2}$. We then have\begin{align}
-\log_2 (p_\mathrm{code}\cdot (1-p_\mathrm{code}))=2.\label{eq:compressionbound}
\end{align}
With $P_Z^{\cal{C}}(\vecc) = 2^{-\ell(\vecc)}$ it follows
\begin{align}
\frac{\kl(P_Y\Vert P_Z^{\cal{C}})}{\expop[\ell(Y)]}&=\frac{\displaystyle\sum_{\vecc\in\supp P_Y}{P_Y(\vecc)\log_2{\frac{P_Y(\vecc)}{P_Z^{\cal{C}}(\vecc)}} }}{\expop[\ell(Y)]}\nonumber\\
&=\frac{\displaystyle\sum_{\vecc\in\supp P_Y}{P_Y(\vecc)\log_2{P_Y(\vecc)}}}{\expop[\ell(Y)]}\nonumber\\
&+\frac{\displaystyle\sum_{\vecc\in\supp P_Y}{P_Y(\vecc)\ell(\vecc)}}{\expop[\ell(Y)]}\nonumber.
\end{align}
For the first term of this sum we get
\begin{align}
\frac{\displaystyle\sum_{\vecc\in\supp P_Y}{P_Y(\vecc)\log_2{P_Y(\vecc)}}}{\expop[\ell(Y)]} = - \frac{\entop(P_Y)}{\expop[\ell(Y)]}\nonumber.
\end{align}
The second term of the sum is equal to one, as 
\begin{align}
\expop[\ell(Y)] = \sum_{\vecc\in\supp P_Y}{P_Y(\vecc)\log_2{P_Y(\vecc)}}\nonumber.
\end{align}
Thus
\begin{align}
\frac{\kl(P_Y\Vert P_Z^{\cal{C}})}{\expop[\ell(Y)]} = 1- \frac{\entop(P_Y)}{\expop[\ell(Y)]}\nonumber
\end{align}
holds. With $\entop(P_Y) = \entop(P_{S^n}) = n\entop(P_S)$ we get
\begin{align}
\expop[\ell(Y)]&=n\entop(P_S)+\kl(P_Y\Vert P_Z^{\cal{C}})\nonumber\\
&\oleq{a} n\entop(P_S)+2\label{eq:compressionlengthbound}
\end{align}
where (a) follows by \eqref{eq:klbound} and \eqref{eq:compressionbound}. The bound \eqref{eq:compressionlengthbound}
recovers the known bound for arithmetic data compression, see \cite[Exercise 6.1]{mackay2003information}. This shows that ADM can be used for arithmetic data compression by using the target distribution $P_Z(0)=P_Z(1)=0.5$. For Huffman codes
\begin{align}
\expop[\ell(Y)] \leq n\entop(P_S) + 1\nonumber
\end{align}
holds \cite[Theorem 5.4.1]{cover2006elements}. The additional bit necessary for arithmetic coding is the price for calculating the codewords online.

\section{Numerical Results}
\label{sec:numerical}

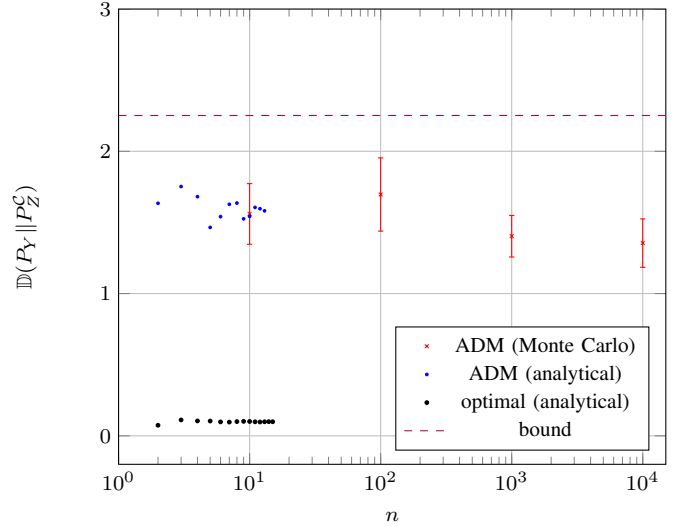
\begin{figure}
\footnotesize
	\centering
	\scalebox{1}{
	\begin{tikzpicture}
	\begin{semilogxaxis}[xlabel =  $n$, width=1\linewidth, grid = major,
	       xmax = 15000,
	       xmin = 1,
               ylabel style = {align=center},
               %ylabel = Wahrscheinlichkeit einer \\ richtigen Dekodierung,
		ylabel = $\kl(P_Y\Vert P_Z^{\cal{C}})$,
	       ymax = 3,
               ymin = -0.2,
		legend pos= south east]
	       %legend pos = north west]

	\addplot[color = red,
		 mark size = 1,
		 mark = x,
		 only marks,
                error bars/.cd,
		 y dir=both,
                y explicit
		] table[x index =0,y index = 1, y error index = 2] {plotdaten2.txt} ;
	\addlegendentry{ADM (Monte Carlo)}
	\addplot[color = blue,
		 mark size = 0.5,
		 mark = *,
		 only marks,
		 %y dir=both,
                 %y explicit
		] table[x index =0,y index = 1] {plotdaten3.txt} ;
	\addlegendentry{ADM (analytical)}
	\addplot[color = black,
		 mark size = 0.7,
		 mark = *,
		 only marks,
		 %y dir=both,
                 %y explicit
		] table[x index =0,y index = 1] {plotdaten4.txt} ;
	\addlegendentry{optimal (analytical)}
	\addplot [purple, dashed, no markers] coordinates {(1,2.2515) (15000,2.2515)};
	\addlegendentry{bound}
	\end{semilogxaxis}
	\end{tikzpicture}}
	\caption{Informational divergence $\kl(P_Y\Vert P_Z^{\cal{C}})$ plotted over input length $n$ for $P_S(0) = 0.5$ and $P_Z(0) = 0.3$.}
	\label{fig: plot}
	\end{figure}
\begin{figure}
\footnotesize
	\centering
	\scalebox{1}{
	\begin{tikzpicture}
	\begin{semilogxaxis}[xlabel =  $n$, width=1\linewidth, grid = major,
	       xmax = 15000,
	       xmin = 1,
               ylabel style = {align=center},
               %ylabel = Wahrscheinlichkeit einer \\ richtigen Dekodierung,
		ylabel = $\frac{\entop(P_Y)}{\expop[\ell(Y)]}$,
	       ymax = 1,
               ymin = 0.5,
		legend pos= south east]
	       %legend pos = north west]

	\addplot[color = red,
		 mark size = 1,
		 mark = x,
		 only marks,
                 error bars/.cd,
		 y dir=both,
                 y explicit
		] table[x index =0,y index = 1, y error index = 2] {plotdatenrateADM.txt} ;
	\addlegendentry{ADM (Monte Carlo)}
	\addplot[color = blue,
		 mark size = 0.5,
		 mark = *,
		 only marks,
		 %y dir=both,
                 %y explicit
		] table[x index =0,y index = 1] {plotdatenrateadmana.txt} ;
	\addlegendentry{ADM (analytical)}
	\addplot[color = black,
		 mark size = 0.7,
		 mark = *,
		 only marks,
		 %y dir=both,
                 %y explicit
		] table[x index =0,y index = 1] {plotdatenrateTunstall.txt} ;
	\addlegendentry{optimal (analytical)}
	\addplot [purple, dashed, no markers] coordinates {(1,0.88129) (15000,0.88129)};
	\addlegendentry{$\entop(P_Z)$}
	\end{semilogxaxis}
	\end{tikzpicture}}
	\caption{Rate $\frac{\entop(P_Y)}{\expop[\ell(Y)]}$ plotted over input length $n$ for $P_S(0) = 0.5$ and $P_Z(0) = 0.3$.}
	\label{fig: plot2}
	\end{figure}
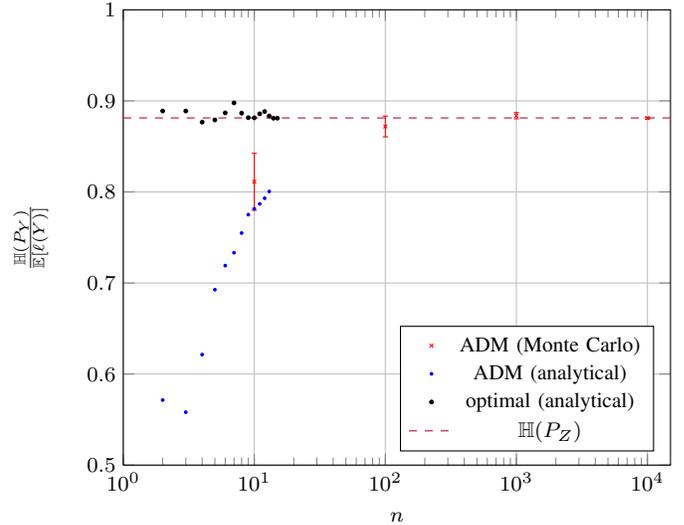
To validate our analytical results in Section~\ref{sec:analysis}, we discuss an example application of our ADM implementation. We consider a uniform binary source $P_S(0)=P_S(1)=0.5$ and the target DMS $P_Z(0)=1-P_Z(1)=0.3$ and we evaluate the informational divergence and the expected output length. For $n=1,2,\dotsc,13$, we calculate the correct values. For $n=10^1,10^2,10^3,10^4$, we use estimates obtained from Monte Carlo simulation. The results for informational divergence are displayed in Figure~\ref{fig: plot}. All obtained values are below the theoretical bound $-\log_2 (0.3\cdot 0.7)=2.2515$. This validates Proposition~\ref{prop:ratiobound} and \eqref{eq:klbound}. In Figure \ref{fig: plot2} we plot the rate $\frac{\entop(P_Y)}{\expop[\ell(Y)]} = \frac{n}{\expop[\ell(Y)]}$ versus the input length $n$. As $n$ gets large, the rate approaches $\entop(P_Z)=0.8813$ from below. This validates Proposition~\ref{prop:rate}.

For comparison, we also calculate informational divergence and rate for the optimal code \cite{amjad2013fixed}. As can be seen in Figure~\ref{fig: plot}, the informational divergence is smaller than for ADM. Suppose now we would like to encode $10^4$ input bits. By \eqref{eq: normalized}, the resulting normalized divergence for ADM would be bounded from above by
\begin{align}
\frac{\kl(P_Y\Vert P_Z^{\cal{C}})}{\expop[\ell(Y)]}&\leq\frac{\log_2\frac{1}{p_\mathrm{code}\cdot (1-p_\mathrm{code})}}{n\entop(P_{S})}=2.2515\cdot 10^{-4}. \nonumber
\end{align}
Alternatively we could apply the optimal code for $n=10$ one thousand times. The resulting normalized divergence is in this case given by
\begin{align}
\frac{\kl(P_Y\Vert P_Z^{\cal{C}})}{\expop[\ell(Y)]}=\frac{1000\cdot 0.1010}{1000\cdot 10/0.8814}=8.9021\cdot 10^{-3} \nonumber
\end{align}
which is higher than for ADM. This shows that using sub-optimal matchers that can encode online is advantageous for large input lengths.

\appendices
\section{Proof of Proposition~\ref{prop:ratiobound}}

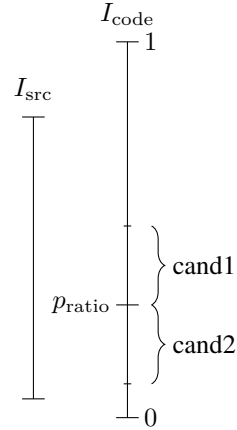
\begin{figure}
\centering
\begin{tikzpicture}[scale = 0.5]

\draw(1,0) -- (1,10);
\draw(0.7,0) -- (1.3,0);
\draw(0.7,10)-- (1.3,10);
\draw(0.7,3) -- (1.3,3);
\node[anchor = east] at (0.8,3) {$p_\mathrm{ratio}$};
\node[anchor = west] at (1.2,0) {$0$};
\node[anchor = west] at (1.2,10) {$1$};
%\node[anchor = west] at (1.2,0.9) {$a$};
%\node[anchor = west] at (1.2,5.1) {$b$};
\draw(0.9,5.1)--(1.1,5.1);
\draw(0.9,0.9)--(1.1,0.9);
%\draw(0.9,1.53)--(1.1,1.53);
%\draw(0.9,3.63)--(1.1,3.63);
%\draw(0,4.9) -- (0,0.7); %wird durch min * 0.5 identifiziert, wenn man das intervall auf eins skaliert, von dem aus man zwei refinements
%braucht um zu finalisieren
%\draw(0.2,4.9)--(-0.2,4.9);
%\draw(0.2,0.7)--(-0.2,0.7);
%\draw(-2,8)--(-2,0.5);%wird durch min * max identifiziert, wenn man das intervall auf eins skaliert, von dem aus man zwei refinements
%braucht um zu finalisieren
%\draw(-1.8,8)--(-2.2,8);
%\draw(-1.8,0.5)--(-2.2,0.5);
\node[anchor = south] at (1,10.2) {$I_\mathrm{code}$};
%\node[anchor = south] at (-2,8) {$I_\mathrm{src}^{(1)}$};
%\node[anchor = south] at (-1,6) {$I_\mathrm{src}^{(2)}$};
%\node[anchor = south] at (0,4.9) {$I_\mathrm{src}^{(3)}$};

%\draw(-0.8,5.9)--(-1.2,5.9);
%\draw(-0.8,1.2)--(-1.2,1.2);
%\draw(-2,8)--(-2,0.5);
%\draw(-1,1.2)--(-1,5.9);

\draw (-1.5,0.5) -- (-1.5,8);
\draw (-1.8,0.5) -- (-1.2,0.5);
\draw (-1.8,8) -- (-1.2,8);
\node[anchor = south] at (-1.5,8.2) {$I_\mathrm{src}$};

\draw [decorate,decoration={brace,mirror,amplitude=5pt},xshift=4pt,yshift=0pt]
(1.5,0.9) -- (1.5,3)node [black,midway,xshift = 20pt] {$\text{cand2}$};
\draw [decorate,decoration={brace,mirror,amplitude=5pt},xshift=4pt,yshift=0pt]
(1.5,3) -- (1.5,5.1)node [black,midway,xshift = 20pt] {$\text{cand1}$};

\end{tikzpicture}
\caption{The scenario for the last step of the algorithm.}
\label{fig: finalize}
\end{figure}

To prove the upper bound on the ratio $I_\mathrm{src}/I_\mathrm{code}$, we have to take a closer look at the last step of the encoding algorithm. We consider the scenario depicted in Figure \ref{fig: finalize}. The last step of the encoding algorithm can always be reduced to this scenario. The encoder refines $I_\mathrm{code}^u$ and $I_\mathrm{code}^l$ independently at the end of the algorithm as described in \ref{sec: encoder}. To achieve the state in Figure \ref{fig: finalize}, we stop each of these refinements one step before a candidate identifies $I_\mathrm{src}$. We then drop all candidates but the two neighboring candidates, where one is a subinterval of $I_\mathrm{code}^u$ and the other is a subinterval of $I_\mathrm{code}^l$. Then we scale this interval consisting of two candidates to $[0,1)$. The last step of the algorithm is now to refine each of the two subintervals of $[0,1)$ and to choose the largest subinterval identifying $I_\mathrm{src}$. The two final candidates are $\mathrm{cand1} = p_\mathrm{code}\cdot(1-p_\mathrm{ratio})$ and $\mathrm{cand2} = p_\mathrm{ratio}\cdot (1-p_\mathrm{code})$. We now show that at least one of the two final candidates is larger than $p_\mathrm{code}\cdot (1-p_\mathrm{code})$. The following statements are equivalent:
\begin{align}
p_\mathrm{ratio}\cdot (1-p_\mathrm{code}) \leq p_\mathrm{code}\cdot (1-p_\mathrm{code})\nonumber\\
\Leftrightarrow p_\mathrm{ratio} \leq p_\mathrm{code}\nonumber\\
\Leftrightarrow (1-p_\mathrm{ratio}) \geq (1-p_\mathrm{code})\nonumber\\
\Leftrightarrow p_\mathrm{code}\cdot (1-p_\mathrm{ratio}) \geq p_\mathrm{code}\cdot (1-p_\mathrm{code}).\nonumber
\end{align}
This shows in particular that if cand2 is smaller than $p_\mathrm{code}\cdot (1-p_\mathrm{code})$ then cand1 is larger than $p_\mathrm{code}\cdot (1-p_\mathrm{code})$. Similarly,
\begin{align}
p_\mathrm{code}\cdot (1-p_\mathrm{ratio}) \leq p_\mathrm{code}\cdot (1-p_\mathrm{code})\nonumber\\
\Leftrightarrow 1-p_\mathrm{ratio} \leq 1-p_\mathrm{code}\nonumber\\
\Leftrightarrow p_\mathrm{ratio} \geq p_\mathrm{code}\nonumber\\
\Leftrightarrow p_\mathrm{ratio}\cdot (1-p_\mathrm{code}) \geq p_\mathrm{code}\cdot (1-p_\mathrm{code})\nonumber
\end{align}
which shows that if cand1 is smaller than $p_\mathrm{code}\cdot (1-p_\mathrm{code})$ then cand2 is larger than $p_\mathrm{code}\cdot (1-p_\mathrm{code})$. Since the algorithm chooses for the final code interval $I_\text{code}^\text{final}=\max\{\mathrm{cand1,cand2}\}$, we have
\begin{align}
\frac{I_\mathrm{src}}{I_\text{code}^\text{final}}=\frac{I_\mathrm{src}}{\max\{\mathrm{cand1,cand2}\}}\leq\frac{1}{p_\mathrm{code}\cdot (1-p_\mathrm{code})}\nonumber
\end{align}
which is the statement of the proposition.

\newpage
\bibliographystyle{IEEEtran}
\normalsize
\bibliography{IEEEabrv,confs-jrnls,references}

\end{document}